\documentclass[10pt,a4paper,twocolumn]{article}
\usepackage[a4paper,left=1.9cm,right=1.9cm,top=2.4cm,bottom=2.4cm,columnsep=0.7cm]{geometry}
\usepackage{amsmath}
\IfFileExists{newtxtext.sty}{\usepackage{newtxtext,newtxmath}}{\usepackage{times,amssymb}}
\usepackage[T1]{fontenc}
\usepackage{graphicx}
\usepackage{bm}
\usepackage{array}
\usepackage{tabularx}
\usepackage{booktabs}
\usepackage{url}
\usepackage{placeins}
\usepackage[authoryear,round]{natbib}
\usepackage[colorlinks=true,linkcolor=red,citecolor=blue,urlcolor=blue]{hyperref}
\usepackage{orcidlink}
\usepackage{caption}
\title{Compression of turbulent flow data through an extended space--time correlation: certification against the modal energy budget}
\author{Gustavo Lopes\textsuperscript{1,*}\,\orcidlink{0009-0009-5061-897X}, Davide Lengani\textsuperscript{1}\,\orcidlink{0000-0001-6347-0817} and Dan Henningson\textsuperscript{2}\,\orcidlink{0000-0001-7864-3071}}
\date{\normalsize \textsuperscript{1}DIME, University of Genova, Genova, Italy\\
\textsuperscript{2}FLOW, Department of Engineering Mechanics, KTH Royal Institute of Technology, Stockholm, Sweden\\[4pt]
\small \textsuperscript{*}Corresponding author: \href{mailto:gustavo.filipe.docarmolopes@edu.unige.it}{gustavo.filipe.docarmolopes@edu.unige.it}}

\begin{document}
\twocolumn[
\begin{@twocolumnfalse}
\maketitle
\vspace{-6pt}
\noindent\rule{\linewidth}{0.4pt}\par\vspace{4pt}
\noindent\textbf{Abstract}\par\vspace{2pt}
\noindent The fidelity of direct numerical simulation now outpaces the storage and data-movement capacity available for the resulting fields. We present an open, distributed, GPU-accelerated low-rank compression framework for large turbulent flow databases. We assess it using two direct numerical simulations of a controlled diffusion compressor cascade under laminar and turbulent inflow conditions. Compression is formulated using a correlation operator over space and time, which is two orders of magnitude larger than that of snapshot POD. Its dominant range is extracted with a randomized range finder based on communication-avoiding hierarchical QR. The compressed databases are evaluated through a hierarchy of increasingly strict diagnostics, from instantaneous reconstruction and second-order statistics to the modal turbulent kinetic energy budget and its nonlinear interscale transfer. The budget terms are far more sensitive to compression than fieldwise or one-point errors. They remain accurate at conservative compression and degrade sharply once the retained energy is relaxed. The near-wall dissipation and the interscale transfer degrade the most and set the admissible compression level. An aggressive reduction preserves every one-point statistic to within a few per cent while corrupting both by an order of magnitude more. The dissipation error is confined to the viscous sublayer and grows with the development of the boundary layer. It is carried by the high-order modes. For both inflows, the linear budget terms scale with the discarded energy to the power $0.8$ and the triadic transfer to a power above one.
\par\vspace{4pt}
\noindent\textbf{Keywords:} data compression, low-rank approximation, turbulent kinetic energy budget, interscale energy transfer, randomized linear algebra, GPU computing
\par\vspace{4pt}\noindent\rule{\linewidth}{0.4pt}
\vspace{10pt}
\end{@twocolumnfalse}
]

\section{Introduction}\label{sec:intro}
\begin{table*}[!t]
\caption{Compression studies of turbulent flow data, on a common footing. Method: L linear (a projection with an exact truncation error), N nonlinear. $N$: number of entries compressed, grid points $\times$ variables $\times$ snapshots, from the dimensions stated in each paper, the per-snapshot input for learned models. Reduction: original size over compressed size as reported by each paper, converted from a fraction (a) or a bit count (b), or the ratio of full to retained mode count where only a decomposition is reported (c). Cost: compression wall time and hardware, or training cost for learned models. Certified: the quantities compared against the reference, F instantaneous field error, S one-point and second-order statistics, E spectra, G gradients, enstrophy or dissipation, B transport budget terms, T interscale transfer, M modal quantities, parentheses for a partial comparison. n.r.\ not reported.}
\label{tab:literature}
\footnotesize
\setlength{\tabcolsep}{3pt}
\renewcommand{\arraystretch}{1.1}
\newcolumntype{Y}{>{\raggedright\arraybackslash}X}
\begin{tabularx}{\textwidth}{@{}>{\raggedright\arraybackslash}p{2.3cm}YYccc>{\raggedright\arraybackslash}p{2.1cm}l@{}}
\toprule
Reference & Method & Flow & Dim.\ & $N$ & Reduction & Cost & Certified\\
\midrule
\citet{Farge1992} & wavelet packets, threshold (N) & 2D decaying turbulence & 2D & $2\times10^{4}$ & 2--200 & Cray-2, n.r. & (F) E (G)\\
\citet{Otero2018} & Legendre truncation, $L^2$ bound (L) & pipe $Re_\tau=180$, wing $Re_c=10^5$ & 3D & $1\times10^{10}$ & 40$^{\rm a}$ & n.r. & (F) S E\\
\citet{Dunton2020} & single-pass randomized SVD, ID (L) & channel $Re_\tau=180$ & 2D, 3D & $1\times10^{9}$ & 150--400 & serial, s--min & F S (M)\\
\citet{Glaws2020} & convolutional autoencoder (N) & HIT $Re_\lambda\le105$, TGV, channel $Re_\tau=180$ & 3D & $6\times10^{6}$ & 64 & train.\ n.r., V100 & F S E G (B)\\
\citet{Jia2021} & wavelet, quantization, max-norm bound (L) & urban LES $Re_H=1.7\times10^4$ & 3D & $3\times10^{12}$ & $\approx100$ & n.r. & F S E (G)\\
\citet{Momenifar2022,Momenifar2022dcc} & vector-quantized autoencoder (N) & HIT $Re_\lambda=90$, DIT, TGV, channel & 3D & $6\times10^{6}$ & 85$^{\rm b}$ & train.\ 8~h, P100 & F S E G\\
\citet{Zhang2023swin} & Swin transformer (N) & HIT $Re_\lambda=418$, channel $Re_\tau\le550$ & 2D & $5\times10^{4}$ & 16--256 & train.\ 40~h, TITAN RTX & F S E (G)\\
\citet{Glazkov2024} & Johnson--Lindenstrauss projection (L) & 2D blade row $Re=10^5$ & 2D & $4\times10^{9}$ & 6.5--32 & s, laptop & M\\
\citet{Fukami2025} & convolutional autoencoder (N) & boundary layer $Re_\tau\le1024$ & 2D & $3\times10^{8}$ & n.r. & n.r. & F (M)\\
\citet{Fung2025} & vector-quantized VAE, images (N) & channel $Re_\tau=395$ & 2D & n.r. & n.r. & n.r. & (F)\\
\citet{Pisoni2026} & quantics tensor train (L) & HIT $Re_\lambda=315$ & 3D & $3\times10^{9}$ & 45$^{\rm a}$ & n.r. & E (G)\\
\citet{Miro2026} & randomized GPU QR--SVD (L) & Windsor body LES $Re_L=2.9\times10^6$ & 3D & $6\times10^{11}$ & 4.2$^{\rm c}$ & 3~s, 100 H100 & M\\
\citet{Perez2026} & Gaussian-process regression (N) & pipe $Re_\tau=180$, channel, wing & 1D, 3D & $7\times10^{7}$ & 5--71 & 100--320~s, MI250x & F S G B\\
\midrule
present & randomized low-rank space--time correlation (L) & compressor cascade $Re=2.5\times10^5$ & 3D & $3\times10^{11}$ & 4.5--28 & 9.6 node-h, A100 & F S G B T M\\
\bottomrule
\end{tabularx}
\end{table*}
The fidelity of direct numerical simulation (DNS) has outgrown the storage and data-movement capacity available for the resulting fields \citep{Dunton2020,Glaws2020,Momenifar2022}. Public turbulence databases hold hundreds of terabytes \citep{Perlman2007,Li2008}. One of \citet{Meneveau2025} reaches petabytes. Lossy compression offers large reductions, provided the discarded information does not compromise the physics of interest. The question is which physical content must be preserved.

Table~\ref{tab:literature} places the compression studies of turbulent flow data on a common footing. The methods span wavelet and orthogonal transforms, matrix and tensor decompositions, Gaussian-process regression and learned models, linear (L) or nonlinear (N). The flows are mostly canonical, isotropic turbulence and channel flow at $Re_\tau\le550$, with one blade row, one wing and one urban LES. The number of entries compressed ranges from $10^{4}$ to $10^{12}$, with the learned models trained on $10^{4}$ to $10^{8}$ entries per snapshot. The reduction factors range from $4$ to $400$. The cost is reported for a minority of the studies, from seconds on a laptop to tens of hours of training on one accelerator. The certification column shows that quality is judged by fieldwise errors (F), one-point statistics (S), and spectra (E), with gradient statistics (G) in about half of the studies. These measures are necessary but not sufficient. A field can reproduce them while corrupting the gradient- and correlation-dependent terms of the energy budget, which are the quantities now examined directly on stored data through the modal interscale transfer formalism \citep{Waleffe1992,Baj2017,Cardesa2017,Cho2018,Ding2025,Nekkanti2025,Bilbao2025}. A transport budget (B) enters the table once, the local time-averaged budget of channel flow of \citet{Perez2026}. Error-bounded tools such as SZ, ZFP, MGARD, and SPERR \citep{Diffenderfer2019,Li2023sperr,Di2025} give a per-sample guarantee that does not transfer to gradient-based or spectral quantities \citep{Laney2013,Ren2026,Cappello2025}. The modal budget and its triadic transfer (T) have not served as validation tools for compression. Read together, the reduction and certification columns show that the learned models reach the largest factors on three-dimensional turbulence, $64$ to $85$, at the cost of the smallest scales. The autoencoder of \citet{Glaws2020} ``filters out the small-scale features'' and the gradient statistics of \citet{Momenifar2022} degrade at the finest scales. Their decoder is a nonlinear map, so the reconstruction is not a projection and carries no Eckart--Young statement about what was discarded. A linear low-rank reduction is a projection with an exact truncation error, needs no training, and stores a modal basis of the data. Reduction factors are therefore not comparable across fidelity levels. The comparison that settles the question is the one made here: the same hierarchy of diagnostics up to the triadic transfer applied to the compressed field.

This work presents an open, distributed, GPU-accelerated framework for the lossy compression of large turbulent flow databases. It is built on a randomized, communication-avoiding hierarchical QR reduction of the correlation operator \citep{Demmel2012,Halko2011} and is archived on Zenodo at \url{https://doi.org/10.5281/zenodo.22013927}. The framework is applied to two DNS databases of a controlled diffusion airfoil cascade, at laminar and turbulent inflow. The contribution is fourfold. (i) An open, distributed compression workflow built on a correlation operator far larger than that of snapshot POD. (ii) A fidelity assessment up to the modal budget and its triadic transfer. (iii) The dependence of this fidelity on inflow condition and retained energy, resolved along the wall, across the boundary layer, and over the mode hierarchy. (iv) A power-law scaling of every budget error with the discarded energy. The framework targets the a posteriori compression of archived databases, where fidelity rather than throughput is the constraint, not in situ compression concurrent with the solver \citep{Pacella2022}.

The paper is structured as follows. Section~\ref{sec:case} defines the flow cases. Section~\ref{sec:method} presents the data matrix, the correlation-operator formulation, the randomized reduction, and the storage. Section~\ref{sec:impl} describes the implementation and its cost. Section~\ref{sec:diag} defines the diagnostics. Section~\ref{sec:results} reports the results. Section~\ref{sec:concl} draws the conclusions.

\section{Flow cases and reference data}\label{sec:case}
\begin{figure}
\centering
\includegraphics[width=\columnwidth]{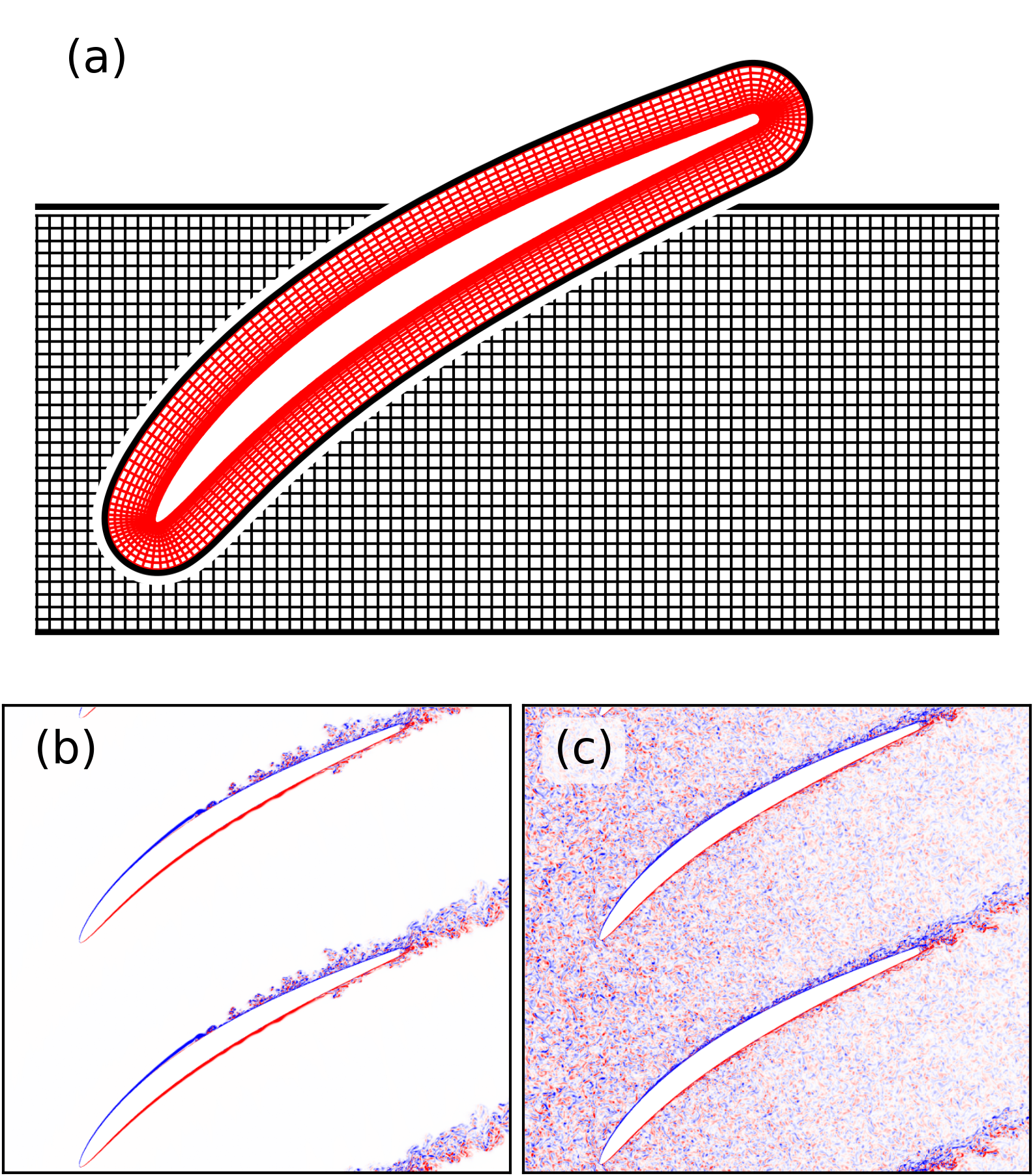}
\caption{Reference flow cases and computational set-up. (a) Overset domain, with the Cartesian background mesh and the O-grid around the controlled diffusion airfoil. Only the near-wall O-grid block (red) is retained. (b, c) Instantaneous spanwise vorticity for the laminar and turbulent inflow cases.}
\label{fig:setup}
\end{figure}

The framework is assessed on a controlled diffusion airfoil (\emph{CDA}) cascade representative of a modern high-pressure compressor \citep{Przytarski2021}. The reference fields are three-dimensional DNS computed with the high-order compressible solver HiPSTAR, validated for turbomachinery applications \citep{Sandberg2015}. The simulations use an overset domain combining an O-grid around the blade with a Cartesian background mesh \citep{Deuse2020}, shown in figure~\ref{fig:setup}a. Two reference databases are considered, at the same inlet Reynolds and Mach numbers, $Re\approx2.5\times10^{5}$ and $Ma=0.20$, differing only in inflow disturbance level: $Tu\approx0.063\%$ for the laminar and $Tu\approx7.9\%$ for the turbulent inflow. The former develops turbulence within the passage, following a laminar separation bubble on the suction side (figure~\ref{fig:setup}b). The latter carries broadband fluctuations already at the inlet (figure~\ref{fig:setup}c). Synthetic inflow turbulence for the turbulent case is generated with a compressible extension of the digital filter method of \citet{Klein2003}.

\begin{figure}
\centering
\includegraphics[width=\columnwidth]{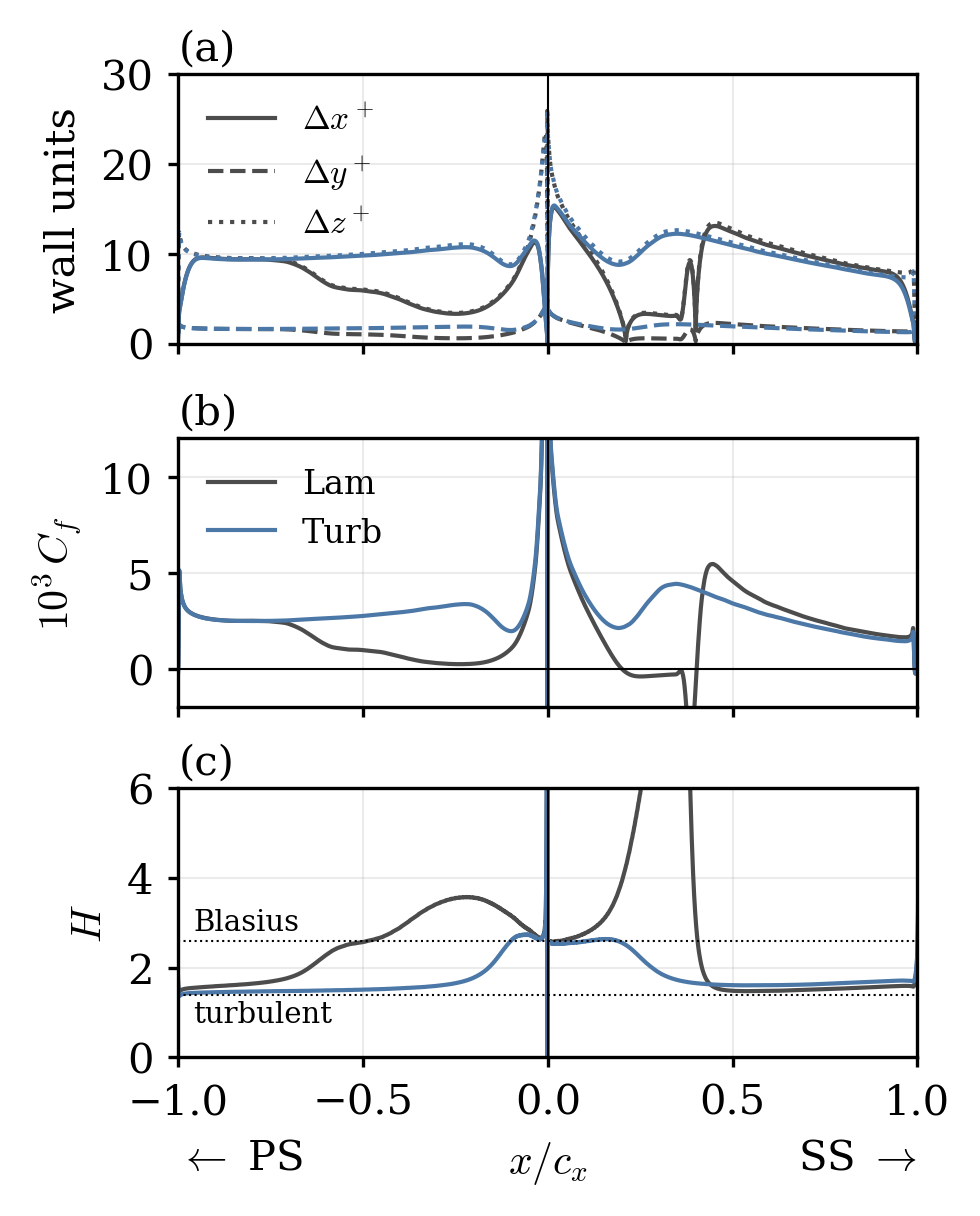}
\caption{Blade boundary layers of the two reference cases along the axial chord, pressure side (PS) on the negative abscissa and suction side (SS) on the positive one. (a) Grid spacings in wall units. (b) Skin-friction coefficient, sign reversed on the pressure side. (c) Shape factor, with the Blasius and turbulent flat-plate values dotted.}
\label{fig:case}
\end{figure}

Only the near-wall O-grid block around the blade is retained (red domain in figure~\ref{fig:setup}a), where the boundary layer and near-wake dynamics relevant to the budget diagnostics live. The coordinates $x$, $y$ and $z$ denote the streamwise, wall-normal and spanwise directions of the O-grid. The block contains $n_x\times n_y\times n_z=3000\times200\times256$ points in these directions, about $1.5\times10^{8}$ in total. Each case retains $n_t=561$ snapshots. Both inflow conditions share the same mesh. Figure~\ref{fig:case}a gives the grid spacings in wall units along the blade for both cases, the pressure side on the negative abscissa. The peak at the leading edge is the stagnation region, where the wall shear and the local wall unit lose their meaning. Along the development of the boundary layers, including the transition region of the laminar case, $\Delta x^{+}$ and $\Delta z^{+}$ stay below $15$ and $\Delta y^{+}$ between $1$ and $2$ for both cases, reaching $2.5$ only close to the leading edge, so both simulations are wall resolved. This is the resolution at which the cascade DNS of \citet{Przytarski2021} was validated against experiment.

The skin-friction coefficient and shape factor of figure~\ref{fig:case}b, c fix the state of the boundary layers that the compression must represent. With laminar inflow, the suction-side boundary layer separates at $x/c_x\approx0.20$ and reattaches at $x/c_x\approx0.40$. The shape factor rises far above the Blasius value inside the bubble and settles to the turbulent value of about $1.4$ downstream of reattachment. On the pressure side, the laminar layer approaches separation near $x/c_x\approx-0.25$, where $C_f$ falls close to zero and $H$ reaches $3.6$. It transitions downstream of it. With turbulent inflow, both sides are attached. The suction side transitions over $0.1\lesssim x/c_x\lesssim0.3$, where $H$ falls from $2.6$ to $1.5$. The two databases therefore place the same mesh and the same number of snapshots in front of a separation-driven transition and of a bypass transition under broadband forcing.

Lengths are normalized by the axial chord $c_x$ and velocities by the inlet velocity $U$, so all quantities reported below are dimensionless. The three fluctuating velocity components of one database, $3\,n_xn_yn_zn_t$ values, occupy $1.03$~TB in single precision. 

\section{Low-rank compression and reconstruction}\label{sec:method}

\subsection{Data matrix}\label{sec:arr}
\begin{figure}
\centering
\includegraphics[width=\columnwidth]{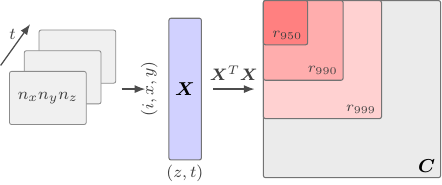}
\caption{Sketch, not to scale, of the data matrix $\bm{X}$ of eq.~\eqref{eq:arrangements}, the snapshots stacked in time and the component index on the rows, and of the correlation operator $\bm{C}=\bm{X}^{T}\bm{X}$ with the three retained ranks of this work nested, $r_{950}<r_{990}<r_{999}$. The side of a rank square over the side of $\bm{C}$ is the storage ratio $r_s\approx r/N_h$.}
\label{fig:arrangements}
\end{figure}

How the data enter the factorization is a choice of the formulation, not a property of the data \citep{Frame2023}. Snapshot POD \citep{Sirovich1987} places one time instant per column of the data matrix. The present formulation places the in-plane locations on the rows and the spanwise direction and time on the columns (figure~\ref{fig:arrangements}). Let $n_x$ and $n_y$ be the numbers of grid points in the two in-plane directions, $n_z$ in the spanwise direction, and $n_t$ the number of snapshots. The three velocity components are stacked into a single data matrix $\bm{X}\in\mathbb{R}^{N_s\times N_h}$, with the component index on the rows, so that one column is a full three-component sample. The row mean of $\bm{X}$ is subtracted from its columns, which for this arrangement is the average over the spanwise direction and time at each in-plane location, $u_i'(\bm{x},t)=u_i(\bm{x},t)-\overline{u}_i(x,y)$. The spanwise direction is homogeneous in the present cases, so this mean is the statistical one, but the algorithm does not require it. The matrix reads
\begin{equation}
X_{\,(i,a,b),\,(c,n)} = u'_{i,abcn},\quad N_s = 3\,n_x n_y,\quad N_h = n_z n_t,
\label{eq:arrangements}
\end{equation}
where $a$, $b$, $c$ and $n$ are the indices along $x$, $y$, $z$ and $t$, $i$ is the velocity component, and the bracketed tuples are row-major compound indices. The rows are the in-plane locations. The columns are the spanwise points over time, $N_h=n_zn_t\approx1.4\times10^{5}$ against $n_t=561$ for snapshot POD. The columns define the objects the truncated basis must represent. The rows define the ensemble over which their correlation is measured \citep{Frame2023}. The correlation operator is therefore two orders of magnitude larger than that of snapshot POD. The truncation acts on the whole volume at once.

\subsection{Correlation operator}\label{sec:corrop}
Rather than factorizing $\bm{X}$, the compression is formulated through the correlation operator
\begin{equation}
\bm{C}=\bm{X}^{T}\bm{X}\in\mathbb{R}^{N_h\times N_h},
\label{eq:corr}
\end{equation}
assembled blockwise from chunks of $\bm{X}$. The dominant matrix products are distributed over many accelerators. The dense reduced solve stays at the size of the column index. This operator differs from the correlation matrix of snapshot POD, whose size is the number of snapshots. Here the column space couples the spanwise direction to time, $N_h\approx1.4\times10^{5}$ against $n_t=561$, too large for a dense eigendecomposition on one device. The decomposition is cast as a low-rank approximation of this operator, whose dynamically relevant range is much smaller. The eigenvectors $\bm{v}_i$ and eigenvalues $\lambda_i=\sigma_i^{2}(\bm{X})$ of $\bm{C}$ are the right singular vectors and squared singular values of $\bm{X}$. Truncation to the leading $r$ eigenpairs is the Frobenius-optimal rank-$r$ approximation, with exact error $\|\bm{X}-\bm{X}_r\|_F^{2}=\sum_{i>r}\lambda_i$.

The Gram route squares the condition number, $\mathrm{cond}_2(\bm{C})=\mathrm{cond}_2(\bm{X})^{2}$ \citep{Higham2002}. Eigenvalues below $\varepsilon_{\mathrm{mach}}\lambda_1$ are lost, so singular values of $\bm{X}$ below about $\sqrt{\varepsilon_{\mathrm{mach}}}\,\sigma_1$ cannot be resolved. The tail of the retained basis sits at the single-precision floor of the Gram products. The consequence is measured by the verification test of section~\ref{sec:verify}.

\subsection{Randomized reduction}\label{sec:randred}
\begin{figure*}[!t]
\centering
\includegraphics[width=\textwidth]{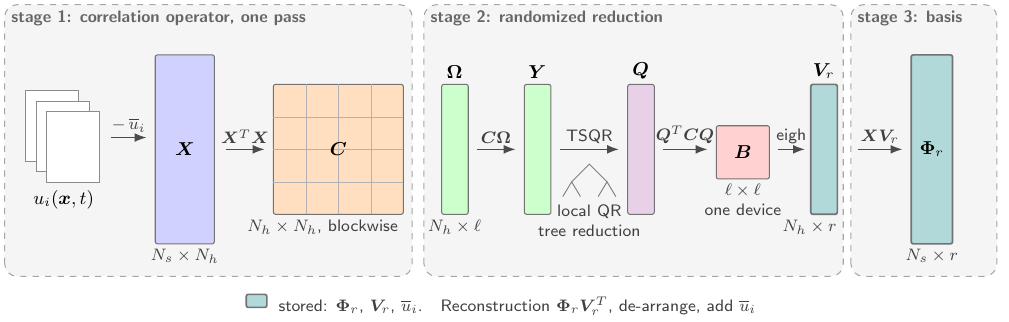}
\caption{Compression pipeline: snapshot ingestion and blockwise contraction to $\bm{C}$ (stage 1), randomized sketch, hierarchical TSQR and reduced eigendecomposition (stage 2), lift and formation of $\bm{\Phi}_r$ (stage 3). Notation of section~\ref{sec:method}. The last two arrays, $\bm{V}_r$ and $\bm{\Phi}_r$, are the stored ones.}
\label{fig:pipeline}
\end{figure*}

The dominant range of $\bm{C}$ is approximated by the randomized sketch $\bm{Y}=\bm{C}\bm{\Omega}$, where $\bm{\Omega}$ is a Gaussian test matrix with $\ell=r+p$ columns \citep{Halko2011}. The sketch is orthonormalized through a blockwise TSQR factorization (figure~\ref{fig:pipeline}). Local QR factorizations are computed on each accelerator. The triangular factors are hierarchically reduced and propagated back to assemble an orthonormal basis $\bm{Q}$, which limits communication between accelerators \citep{Demmel2012,Halko2011}. Distributed and randomized variants of this kernel are an active area of research on current GPU systems \citep{Fukaya2020,Mijic2025,Higgins2025}. The reduced operator $\bm{B}=\bm{Q}^{T}\bm{C}\bm{Q}\in\mathbb{R}^{\ell\times\ell}$ is diagonalized on a single accelerator, $\bm{B}=\bm{W}\bm{\Lambda}\bm{W}^{T}$. Its eigenvectors are lifted through $\bm{Q}$ to the compression basis $\bm{V}_r=\bm{Q}\bm{W}_r$, retaining $r$ modes at a prescribed energy target. The oversampling $p$ controls the expected approximation error of the range finder, which is bounded by the tail of the spectrum of $\bm{C}$ \citep{Halko2011}, so the randomized truncation inherits the decay of the tail spectrum. All runs use $p=30$.

The retained energy
\begin{equation}
E_r=\sum_{i=1}^{r}\lambda_i\Big/\sum_{i}\lambda_i
\label{eq:Er}
\end{equation}
is the cumulative fraction of the eigenvalues $\lambda_i$ of $\bm{C}$. Because $\bm{X}$ holds the velocity fluctuations, it is the fraction of the turbulent kinetic energy captured by the truncation, grid-point weighted rather than volume weighted. The O-grid clustering therefore overweights the near-wall region relative to a volume integral. The spatial structures $\bm{\Phi}_r=\bm{X}\bm{V}_r$ follow by one product.

\subsection{Storage and reconstruction}\label{sec:storage}
Only $\bm{\Phi}_r\in\mathbb{R}^{N_s\times r}$ and $\bm{V}_r\in\mathbb{R}^{N_h\times r}$ are stored, with the mean field, all in single precision. The stored size and the storage ratio relative to $\bm{X}$ are
\begin{equation}
M_r=4\,r\,(N_s+N_h)\ \mathrm{B},\ \ r_s=\frac{M_r}{M_0}=r\,\frac{N_s+N_h}{N_sN_h}\approx\frac{r}{N_h},
\label{eq:storage}
\end{equation}
with $M_0=4N_sN_h$ the uncompressed fluctuation data. The achievable reduction is governed by the number of columns of $\bm{C}$ relative to the rank needed to reach the energy target. The field is reconstructed as $\bm{\Phi}_r\bm{V}_r^{T}$ and de-arranged. The mean is restored.

\section{Distributed GPU implementation and cost}\label{sec:impl}
The framework is implemented in Python on Dask and CuPy. The pipeline has three stages (figure~\ref{fig:pipeline}). Stage 1 assembles $\bm{C}$ in one pass over the raw snapshots, with mean subtraction and blockwise contraction on the accelerators. Stage 2 is the randomized eigendecomposition of $\bm{C}$. Stage 3 forms and stores $\bm{\Phi}_r$. Fields stay distributed across the workers. Every large product is computed where its operands reside. Only the reduced objects are gathered. Table~\ref{tab:alg} lists the workflow with the library operations as executed.

\begin{table*}
\caption{Distributed GPU workflow of the low-rank compression, with the library operations as executed. Input: snapshot files, data matrix of eq.~\eqref{eq:arrangements}, energy target or rank $r$, oversampling $p$. Output: $\bm{\Phi}_r$, $\bm{V}_r$, $\overline{u}_i$. Reconstruction $\bm{\Phi}_r\bm{V}_r^{T}$, mean restored.}
\label{tab:alg}
\centering
\small
\begin{tabularx}{\textwidth}{@{}X@{}}
\toprule
\textit{Stage 1: correlation operator, one pass over the data} \\
1. read snapshots, subtract $\overline{u}_i$, arrange into $\bm{X}$ ($N_s\times N_h$), lazy \texttt{dask.array} persisted across workers \\
2. $\bm{C}\gets\bm{X}^{T}\bm{X}$ ($N_h\times N_h$), blockwise tiled \texttt{einsum} on the accelerators, written to disk \\
\textit{Stage 2: randomized reduced eigendecomposition} \\
3. $\bm{\Omega}\gets\mathcal{N}(0,1)$, Gaussian test matrix, $N_h\times(r+p)$ \\
4. $\bm{Y}\gets\bm{C}\bm{\Omega}$, tiled GPU \texttt{einsum} \\
5. $\bm{Q}\gets\mathrm{TSQR}(\bm{Y})$, local QR per row block, tree reduction of the triangular factors \\
6. $\bm{B}\gets\bm{Q}^{T}\bm{C}\bm{Q}$, streamed tiled GPU path, in-place data-local reduction, $(r+p)\times(r+p)$ \\
7. $(\bm{\Lambda},\bm{W})\gets\texttt{eigh}(\bm{B})$, dense symmetric eigensolver on one accelerator, eigenpairs in descending order, truncated at the energy target \\
8. $\bm{V}_r\gets\bm{Q}\,\bm{W}_{[:,1:r]}$ ($N_h\times r$), distributed product \\
\textit{Stage 3: basis formation and storage} \\
9. $\bm{\Phi}_r\gets\bm{X}\bm{V}_r$ ($N_s\times r$), row strips of $\bm{X}$ against column strips of $\bm{V}_r$ \\
10. store $\{\bm{\Phi}_r,\bm{V}_r,\overline{u}_i\}$ in single precision \\
\bottomrule
\end{tabularx}
\end{table*}

All runs use the Booster partition of the Leonardo supercomputer at CINECA. Each node carries one Intel Xeon Platinum 8358 processor with 32 cores, 512~GB of memory, four NVIDIA A100 accelerators with 64~GB each, and a dual-rail NVIDIA HDR InfiniBand network. The pipeline runs one Dask worker per accelerator. The reduced problem $\bm{B}$ is dense and symmetric of order $\ell$. The vendor eigensolver bounds it at about $32{,}000$ modes on one accelerator, which caps the retained rank of the most conservative turbulent database (section~\ref{sec:results:perf}). The complete compression of one database required $9.6$ node-hours, or $38.4$ GPU-hours, covering reading, correlation, the reduced eigendecomposition, and reconstruction.

For comparison, the reference simulation advances on $7$ nodes of the same partition at $0.25$~s per time step. Snapshots are stored every $500$ steps, so the window of $561$ snapshots corresponds to $2.8\times10^{5}$ steps and about $140$ node-hours. The compression is a one-off operation that costs $7\%$ of this window. Its first stage reads the full dataset once and is limited by memory bandwidth on the accelerators.

\label{sec:verify}The assessment of section~\ref{sec:results} assumes that the reconstruction is the orthogonal projection of rank $r$ defined by $\bm{V}_r$. The retained energy does not verify this. $E_r$ is computed from $\operatorname{tr}\bm{C}$ and the eigenvalues of $\bm{B}$, and neither depends on the order in which the row blocks of $\bm{V}_r$ are written or read. An independent validation test, which does not rely on the distributed implementation, is therefore included in the released framework. For one snapshot $t$, it rebuilds the rows $\bm{X}_t$ from the raw file and reads $\bm{V}_r$ once. It then computes three quantities. The first is the orthogonality defect $\max|\bm{V}_r^{T}\bm{V}_r-\bm{I}|$ of the leading columns. The second is the relative difference between the stored rows $\bm{\Phi}_t$ and the recomputed product $\bm{X}_t\bm{V}_r$, which compares the output of stage 3 with stages 1 and 2 without any reconstruction code. The third is the projection error $\|\bm{X}_t-(\bm{X}_t\bm{V}_r)\bm{V}_r^{T}\|^{2}/\|\bm{X}_t\|^{2}$, compared with the truncation value $1-E_r$. Section~\ref{sec:results:field} reports this projection error for every snapshot of the six databases.

\section{Diagnostics}\label{sec:diag}

\subsection{Modal turbulent kinetic energy budget}\label{sec:budget}
The compressed databases are assessed through increasingly strict diagnostics, up to the modal budget of the turbulent kinetic energy (TKE). The fluctuating velocity is decomposed as
\begin{equation}
u_i'(\bm{x},t)=\sum_{\ell=1}^{N_m}u_i^{\ell}(\bm{x},t),
\label{eq:decomp}
\end{equation}
where $u_i^{\ell}$ is the $\ell$th modal contribution and $N_m=n_t$, with modal TKE $k^{\ell}=\tfrac12u_i^{\ell}u_i^{\ell}$. The transport equation for the modal TKE can be written as
\begin{equation}
\mathcal{L}^{\ell}+\mathcal{C}^{\ell}=\mathcal{P}^{\ell}+\mathcal{T}^{\ell}-\varepsilon^{\ell}+\mathcal{D}^{\ell},
\label{eq:budget}
\end{equation}
where $\mathcal{L}^{\ell}$, $\mathcal{C}^{\ell}$, $\mathcal{P}^{\ell}$, $\mathcal{T}^{\ell}$, $\varepsilon^{\ell}$ and $\mathcal{D}^{\ell}$ are the local unsteadiness, convection, production, net interscale transfer, dissipation and grouped diffusion of mode $\ell$. The diffusion comprises pressure transport, turbulent transport and viscous diffusion, $\mathcal{D}^{\ell}=\Pi^{\ell}+\mathcal{T}_{\mathrm{tr}}^{\ell}+\mathcal{D}_\nu^{\ell}$. The modal decomposition follows \citet{Baj2017}, in particular for the triadic interscale transfer $\mathcal{T}^{\ell}$. At $Ma=0.20$ the flow is effectively incompressible and the pressure transport $\Pi^{\ell}$ is small. \citet{Przytarski2024} found pressure-related transport negligible in the mechanical energy budget of a comparable compressor, so $\Pi^{\ell}$ is omitted. The turbulent transport $\mathcal{T}_{\mathrm{tr}}^{\ell}$ shares its third-order moments and spatial triple products with $\mathcal{T}^{\ell}$ and integrates to near zero as a pure spatial redistribution. The sensitivity of the triple correlations is therefore assessed through $\mathcal{T}^{\ell}$ and the diffusion is represented by its viscous part $\mathcal{D}_\nu^{\ell}$. For the statistically stationary databases, the time-averaged $\mathcal{L}^{\ell}$ vanishes.

\subsection{Modal representation of the budget terms}\label{sec:modal}
The reference and reconstructed fluctuating fields are projected onto a snapshot POD basis \citep{Lumley1970,Sirovich1987}, $u_i^{\ell}(\bm{x},t)=a^{\ell}(t)\,\phi_i^{\ell}(\bm{x})$, computed independently of the compression operator $\bm{C}$. The snapshot POD acts on the matrix of size $3n_xn_yn_z\times n_t$, whereas the compression factorizes $\bm{X}$ of size $3n_xn_y\times n_zn_t$, so $N_m=n_t$ modes are obtained. POD is adopted over SPOD \citep{Towne2018} or a Fourier basis because its real-valued, energy-ranked modes make the budget interpretable mode by mode and keep the triadic products free of complex-conjugate bookkeeping. The choice of basis changes the representation of the couplings, not the redistribution they accomplish. The cumulative modal transfer evaluated on POD, SPOD, and Fourier bases of the same record nearly collapses, the SPOD trough being about $5\%$ shallower \citep[Appendix B]{Lopes2026gpu}. The spatial modes $\phi_i^{\ell}$ carry the modal TKE $k^{\ell}=\tfrac12\phi_i^{\ell}\phi_i^{\ell}$. The temporal coefficients $a^{\ell}$ are orthonormal in time. Time enters the budget of mode $\ell$ only through the moments of the coefficients. The single-mode (linear) terms carry $\overline{a^{\ell}a^{\ell}}=1$ and reduce to spatial expressions. The triadic terms carry the third-order moment $\overline{a^{\ell}a^{m}a^{n}}$, evaluated with a GPU-accelerated framework \citep{Lopes2026gpu}. Every term below is therefore the time-averaged contribution of one mode, nonzero for the transfer as well.

The single-mode terms, convection $\mathcal{C}^{\ell}$, production $\mathcal{P}^{\ell}$, dissipation $\varepsilon^{\ell}$ and viscous diffusion $\mathcal{D}_\nu^{\ell}$, together with the net interscale transfer $\mathcal{T}^{\ell}$ read
\begin{equation}
\begin{aligned}
\mathcal{C}^{\ell}&=\overline{u}_j\,\partial_j\bigl(\tfrac12\phi_i^{\ell}\phi_i^{\ell}\bigr),\qquad
\mathcal{P}^{\ell}=-\phi_i^{\ell}\phi_j^{\ell}\,\partial_j\overline{u}_i,\\
\varepsilon^{\ell}&=\nu\,\partial_j\phi_i^{\ell}\,\partial_j\phi_i^{\ell},\qquad
\mathcal{D}_\nu^{\ell}=\nu\,\partial_j\partial_j\bigl(\tfrac12\phi_i^{\ell}\phi_i^{\ell}\bigr),\\
\mathcal{T}^{\ell}&=-\sum_{m,n=1}^{N_m}\overline{a^{\ell}a^{m}a^{n}}\;\phi_i^{\ell}\phi_j^{m}\,\partial_j\phi_i^{n},
\end{aligned}
\label{eq:transfer}
\end{equation}
with $\partial_j=\partial/\partial x_j$. Summed over $\ell$, the transfer vanishes by conservation, so it only redistributes energy across the mode hierarchy. The comparison covers the four single-mode terms and the net interscale transfer. All gradients are evaluated with the same curvilinear metrics and stencils on the reference and on every reconstruction. The diagnostics therefore measure the fidelity of the data, not the self-consistency of the truncation. All terms are summed over the grid points of the O-grid block.

Two error measures are used. The relative $L^{2}$ error of a statistic or budget term is computed as the norm of the difference between the compressed and the reference results, divided by the norm of the reference. For the second-order statistics, the norm is taken over the in-plane locations of the span- and time-averaged field. For the budget terms, it is taken over the mode index. The signed cumulative error of a budget term $\mathcal{F}^{j}\in\{\mathcal{C}^{j},\mathcal{P}^{j},\varepsilon^{j},\mathcal{D}_\nu^{j}\}$ is the sum of the difference between the compressed and the reference term up to mode $\ell$, divided by the sum of the absolute values of the reference term over all modes,
\begin{equation}
E_{\mathrm{cum}}(\ell)=\frac{\sum_{j\le\ell}\bigl(\mathcal{F}^{j}_{\mathrm{rec}}-\mathcal{F}^{j}_{\mathrm{ref}}\bigr)}{\sum_{j}\bigl|\mathcal{F}^{j}_{\mathrm{ref}}\bigr|}.
\label{eq:ecum}
\end{equation}
For the transfer, $\mathcal{F}^{j}=\mathcal{T}^{j}$ is divided by the production activity $\sum_j|\mathcal{P}^{j}_{\mathrm{ref}}|$. The two measures are complementary. The $L^{2}$ error ignores the sign of the modal errors. The cumulative error keeps it. A bias of constant sign across the hierarchy accumulates, whereas errors of alternating sign cancel. The last value of the curve is the net signed error of the term.

\section{Results}\label{sec:results}

\subsection{Compression performance and second-order statistics}\label{sec:results:perf}
Each reference database is compressed at three energy targets, $E_r=0.999$, $0.990$ and $0.950$. This gives six compressed databases, labelled by inflow condition and target from Lam999 to Turb950 (table~\ref{tab:cases}). The retained rank $r$ falls sharply as the target is relaxed, most strongly for Turb950. The achieved energy matches the target in all cases but Turb999. There, the retained rank of $3.2\times10^{4}$ is the cap of the reduced eigensolver and corresponds to $E_r=0.994$. The reduction does not rely on the homogeneous direction. In a test on the same cascade with the wall-normal direction in place of the spanwise one in the row ensemble, the rank needed to reach a given retained energy differed by less than $10\%$ from the one reported here. 

\begin{table}
\caption{Summary of the six compressed databases. The uncompressed fluctuation data occupy $1.03$~TB per case. $E_r$ is the achieved retained energy, $r_s\approx r/N_h$ the storage ratio of eq.~\eqref{eq:storage} and the last column the stored size $M_r$.}
\label{tab:cases}
\centering
\footnotesize
\setlength{\tabcolsep}{4pt}
\begin{tabular}{lccccc}
\toprule
Case & Target $E_r$ & $r$ & $E_r$ & $r_s$ & GB\\
\midrule
Lam999 & 0.999 & 31703 & 0.999000 & 0.2207 & 246\\
Lam990 & 0.990 & 12972 & 0.990000 & 0.0903 & 101\\
Lam950 & 0.950 & 5105  & 0.950000 & 0.0355 & 40\\
Turb999 & 0.999 & 32000 & 0.993842 & 0.2228 & 249\\
Turb990 & 0.990 & 25163 & 0.990000 & 0.1752 & 196\\
Turb950 & 0.950 & 10940 & 0.950006 & 0.0762 & 85\\
\bottomrule
\end{tabular}
\end{table}

The left block of table~\ref{tab:errors} reports the relative $L^{2}$ errors of the Reynolds-stress components, evaluated on the span- and time-averaged fields in the blade-to-blade plane. The errors grow with the aggressiveness of the reduction and are largest at the $0.950$ target for both inflows. The stresses involving the spanwise component, $\overline{w'w'}$ and $\overline{v'w'}$, carry the largest relative errors. The spanwise mean velocity is close to zero, so $w'$ is small, and its stresses are small quantities divided by small references. The shear stress $\overline{u'v'}$ is the least sensitive. Even at the $0.950$ target, the second-order statistics stay within a few per cent. This is the level at which previous compression studies certify a reduction as faithful \citep{Glaws2020,Momenifar2022}.

\begin{table*}
\caption{Relative $L^{2}$ error (\%) of the Reynolds-stress components and of the time-averaged modal budget terms (defined in section~\ref{sec:modal}) for the six compressed databases. The same metric applies to every entry: the norm of the compressed-minus-reference quantity normalized by the reference norm. The second-order errors are evaluated on the span- and time-averaged fields in the physical plane. The budget errors are evaluated on the space-integrated modal vectors.}
\label{tab:errors}
\centering
\footnotesize
\setlength{\tabcolsep}{4pt}
\begin{tabular}{lccccccccccc}
\toprule
& \multicolumn{6}{c}{Second-order, physical plane} & \multicolumn{5}{c}{TKE budget, modal}\\
\cmidrule(lr){2-7}\cmidrule(lr){8-12}
& & & & & & & \multicolumn{4}{c}{Single-mode (linear)} & Triadic\\
\cmidrule(lr){8-11}\cmidrule(lr){12-12}
Case & $\overline{u'u'}$ & $\overline{v'v'}$ & $\overline{w'w'}$ & $\overline{u'v'}$ & $\overline{u'w'}$ & $\overline{v'w'}$ & $\mathcal{C}^{\ell}$ & $\mathcal{P}^{\ell}$ & $\varepsilon^{\ell}$ & $\mathcal{D}_\nu^{\ell}$ & $\mathcal{T}^{\ell}$\\
\midrule
Lam999  & 0.05 & 0.08 & 0.10 & 0.03 & 0.09 & 0.13 & 0.33 & 0.02 & 0.81  & 0.40 & 0.22\\
Lam990  & 0.50 & 0.81 & 1.05 & 0.33 & 0.78 & 1.12 & 2.23 & 0.13 & 5.60  & 2.90 & 2.16\\
Lam950  & 2.48 & 4.30 & 5.54 & 1.55 & 3.62 & 5.89 & 8.15 & 0.47 & 17.19 & 9.67 & 16.86\\
Turb999 & 0.39 & 0.51 & 0.70 & 0.13 & 0.39 & 0.41 & 0.62 & 0.16 & 2.89  & 1.95 & 1.55\\
Turb990 & 0.63 & 0.82 & 1.14 & 0.21 & 0.55 & 0.59 & 0.90 & 0.26 & 4.74  & 2.85 & 2.77\\
Turb950 & 3.06 & 4.12 & 5.89 & 1.05 & 2.68 & 2.85 & 5.48 & 1.00 & 16.48 & 9.62 & 18.68\\
\bottomrule
\end{tabular}
\end{table*}

\begin{figure*}[!t]
\centering
\includegraphics[width=\textwidth]{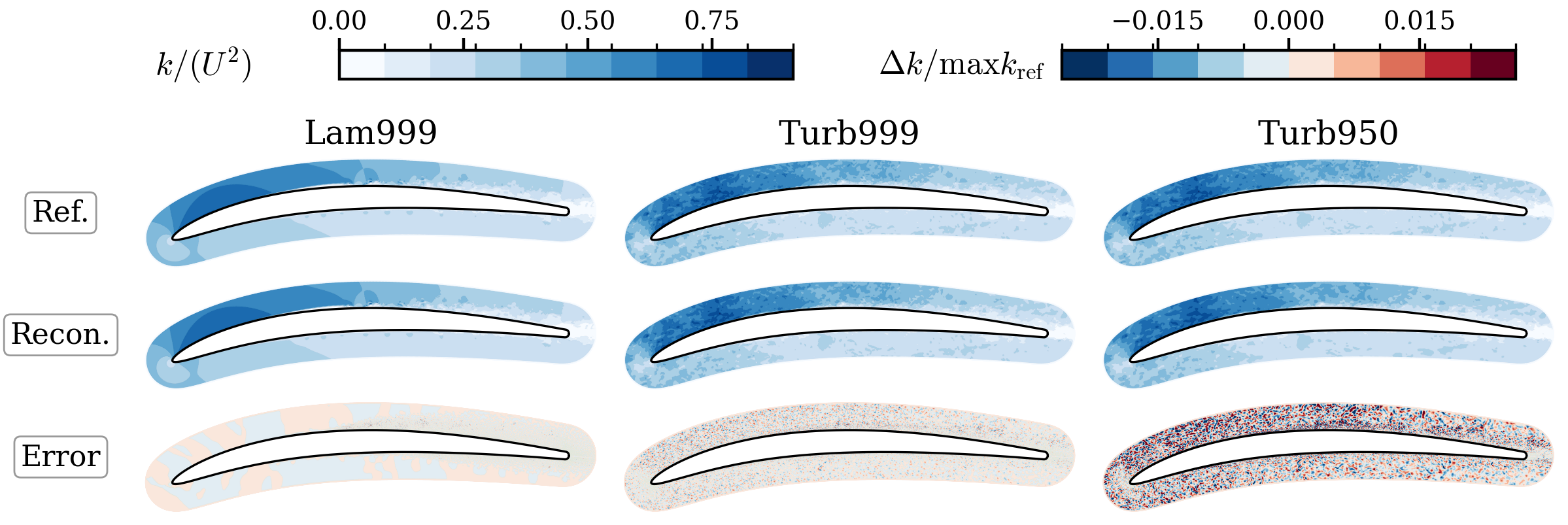}
\caption{Instantaneous fluctuation kinetic energy $k=\tfrac12u_i'u_i'$ on a spanwise plane for Lam999, Turb999 and Turb950 (columns). Top: reference. Centre: reconstruction. Bottom: difference, normalized by the peak reference value in the plane.}
\label{fig:snapshots}
\end{figure*}

\subsection{Instantaneous and local fidelity}\label{sec:results:field}
\begin{figure}
\centering
\includegraphics[width=\columnwidth]{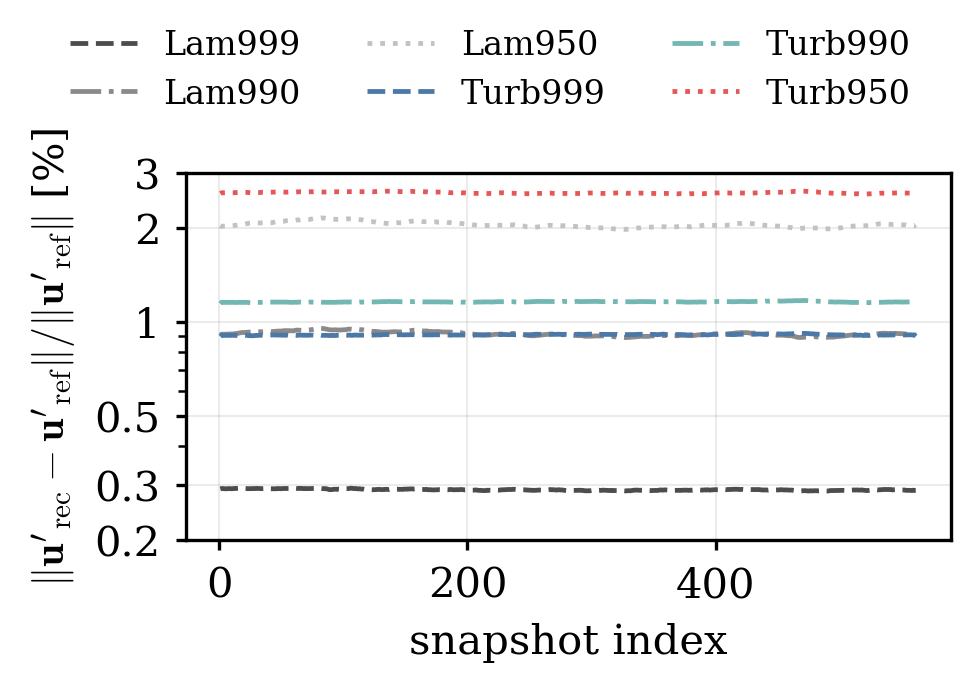}
\caption{Relative $L^{2}$ error of the fluctuating velocity vector of every reconstructed snapshot, for the six compressed databases.}
\label{fig:snaperr}
\end{figure}

Figure~\ref{fig:snapshots} shows the reference and reconstructed instantaneous turbulent kinetic energy $k=\tfrac12u_i'u_i'$ on a spanwise plane for Lam999, Turb999 and Turb950. The peak reference value in the plane normalizes the difference. The conservative reconstructions are indistinguishable from the reference. Turb950 develops a faint broadband footprint over the blade, in the boundary layers and in the turbulent free stream, where the discarded scales live. Figure~\ref{fig:snaperr} quantifies this on every snapshot. The relative $L^{2}$ error of the fluctuating velocity vector is constant in time to within a few per cent of its value. It is $0.29\%$, $0.92\%$ and $2.1\%$ for the laminar inflow and $0.91\%$, $1.2\%$ and $2.6\%$ for the turbulent inflow at the three targets. The truncation does not favour any part of the record. The snapshot-level error of the most aggressive reduction stays below $3\%$.

\begin{figure}
\centering
\includegraphics[width=\columnwidth]{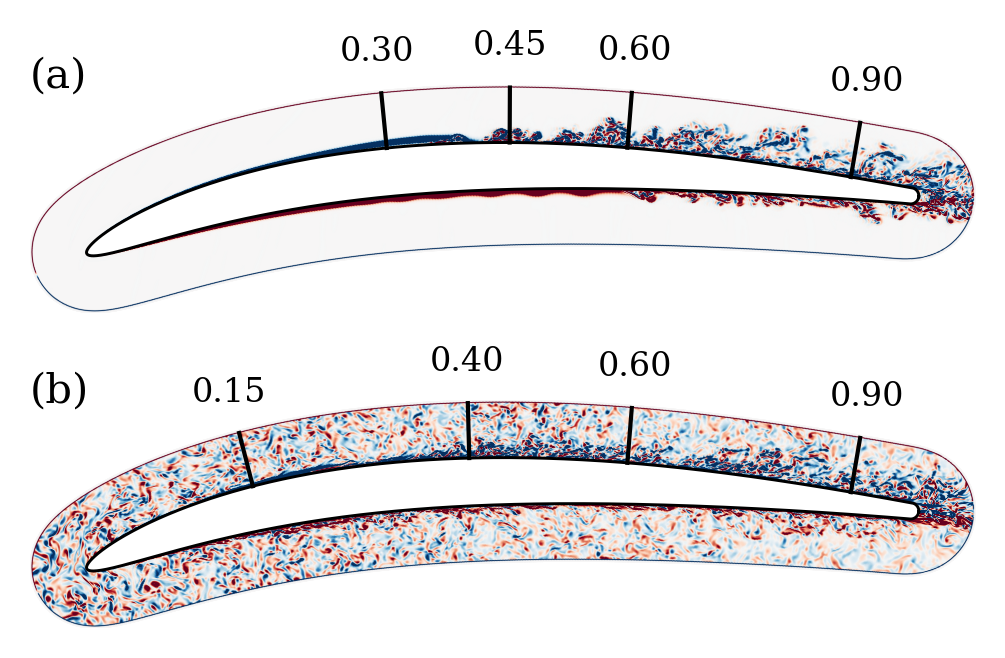}
\caption{Instantaneous spanwise vorticity on a spanwise plane, colour limits clipped, with the suction-side stations of figure~\ref{fig:localdis} marked by their axial position $x/c_x$: (a) laminar inflow, (b) turbulent inflow.}
\label{fig:probes}
\end{figure}

\begin{figure*}
\centering
\includegraphics[width=\textwidth]{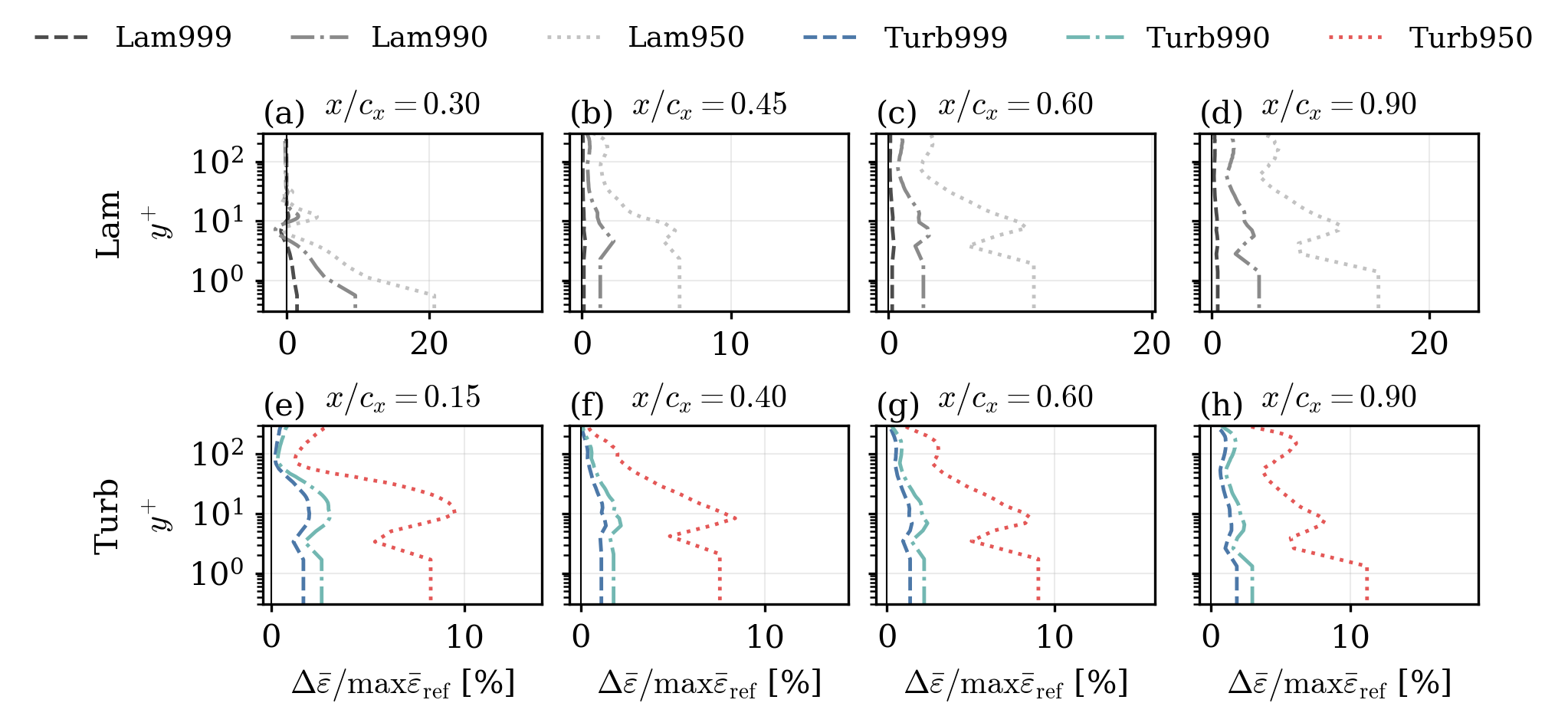}
\caption{Error of the time-averaged modal dissipation, compressed minus reference normalized by the peak reference value of the same profile, against the wall distance in local wall units at the stations of figure~\ref{fig:probes}: (a--d) laminar inflow, $x/c_x=0.30$, $0.45$, $0.60$, $0.90$; (e--h) turbulent inflow, $x/c_x=0.15$, $0.40$, $0.60$, $0.90$. At $x/c_x=0.30$ of the laminar case, the wall shear is reversed and $y^{+}$ is based on $|\tau_w|$. The abscissa scale differs between rows.}
\label{fig:localdis}
\end{figure*}

The hierarchy is then climbed locally, at the wall, where the budget is built on gradients. Figure~\ref{fig:probes} marks four suction-side stations on an instantaneous spanwise vorticity field of each case, chosen from the boundary-layer states of figure~\ref{fig:case}. For the laminar inflow, they are $x/c_x=0.30$ inside the separation bubble, $0.45$ just after reattachment, $0.60$ in the turbulent boundary layer, and $0.90$ near the trailing edge. For the turbulent inflow, they are $0.15$ in the transitional region and $0.40$, $0.60$, and $0.90$ in the attached turbulent layer. Figure~\ref{fig:localdis} reports at each station the wall-normal profile of the error of the dissipation summed over the modes, $\tfrac{1}{N_m}\sum_\ell\varepsilon^{\ell}$, normalized by the peak reference value of the same profile, against the wall distance in local wall units.

Two features are common to every station. The error is largest at the wall and decays across the buffer layer, so the discarded scales are those of the viscous sublayer. It orders with the target at every height. Its magnitude depends on the state of the layer. With laminar inflow, the $0.999$ target holds the dissipation within $1\%$ to $2\%$ at the attached stations and the $0.990$ target within $5\%$ to $8\%$. The $0.950$ target reaches $17\%$ just after reattachment, $19\%$ at $x/c_x=0.60,$ and $23\%$ near the trailing edge. The error grows downstream as the turbulent layer develops. Inside the bubble, at $x/c_x=0.30$, the reference dissipation is almost four orders of magnitude smaller than downstream. The same truncation removes a larger share of it, $16\%$ at the $0.990$ and $34\%$ at the $0.950$ target. With turbulent inflow, the wall error of the $0.950$ target grows from $13\%$ to $18\%$ along the four stations, with a second maximum of $10\%$ at $y^{+}\approx10$ in the transitional station. The conservative targets stay within $3\%$ and $6\%$. 

\subsection{Modal budget fidelity}\label{sec:results:budget}
\begin{figure}
\centering
\includegraphics[width=\columnwidth]{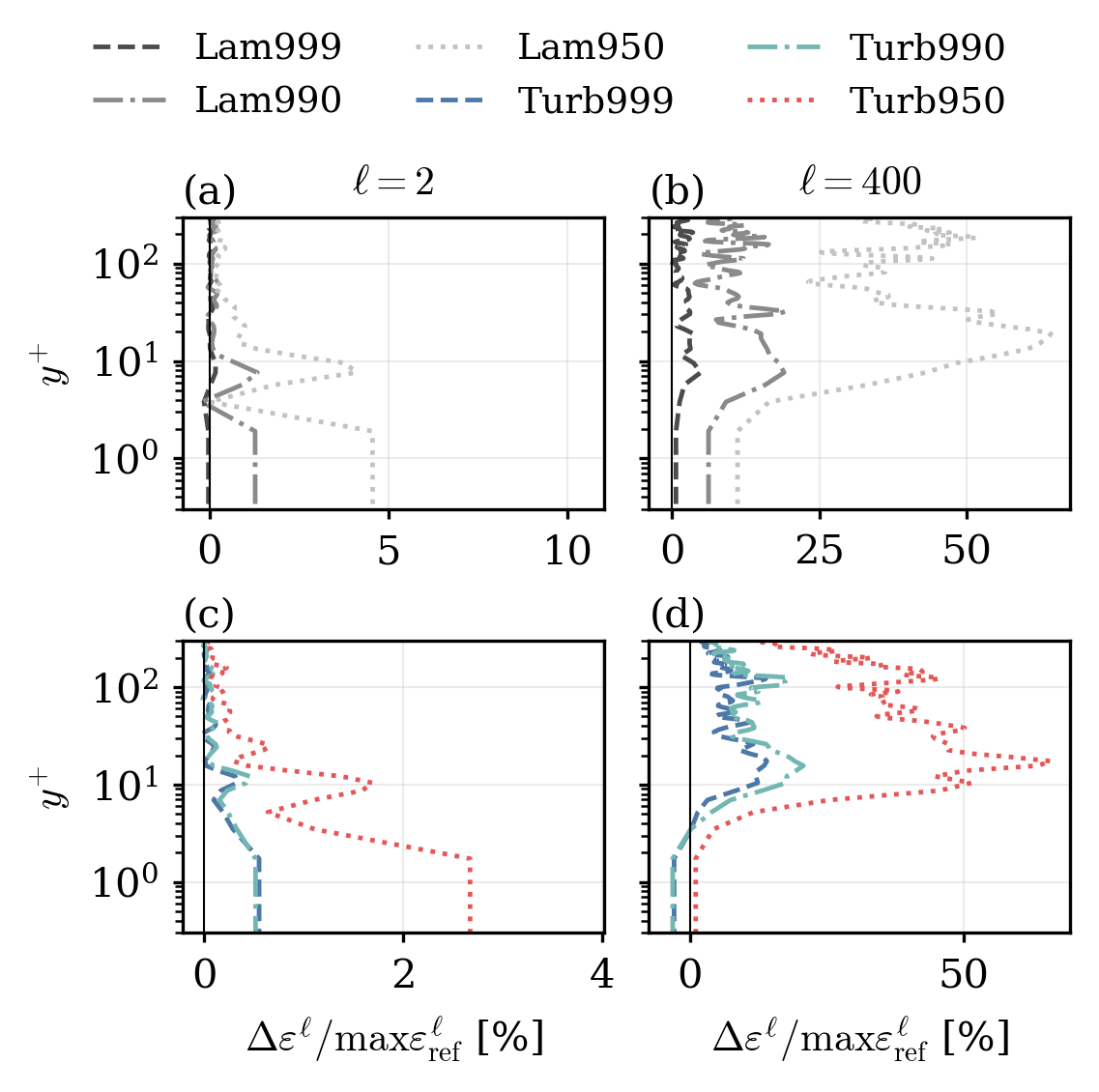}
\caption{Local error of the modal dissipation at $x/c_x=0.60$, compressed minus reference normalized by the peak reference value of the same mode, for the low-order mode $\ell=2$ (a, c) and the high-order mode $\ell=400$ (b, d): (a, b) laminar inflow, (c, d) turbulent inflow. The abscissa scale differs between columns.}
\label{fig:modal}
\end{figure}

The dissipation error of figure~\ref{fig:localdis} is now resolved by snapshot POD mode. Figure~\ref{fig:modal} shows it at the station shared by the two cases, $x/c_x=0.60$, for a low-order mode $\ell=2$ and a high-order mode $\ell=400$. Both modes retain the same temporal coefficients between the reference and the reconstructions, so the same physical modes are compared at every target. The low-order mode is held within $1\%$ across the boundary layer by every conservative target. It reaches $4\%$ to $10\%$ only in the buffer region at the $0.950$ target. The high-order mode is corrupted across the entire boundary layer, by $14\%$ to $21\%$ already at the $0.990$ target and by $64\%$ to $66\%$ at the most aggressive one, in both cases alike. The retained energy protects the low-order modes and leaves the high-order modes exposed.

\begin{figure}
\centering
\includegraphics[width=\columnwidth]{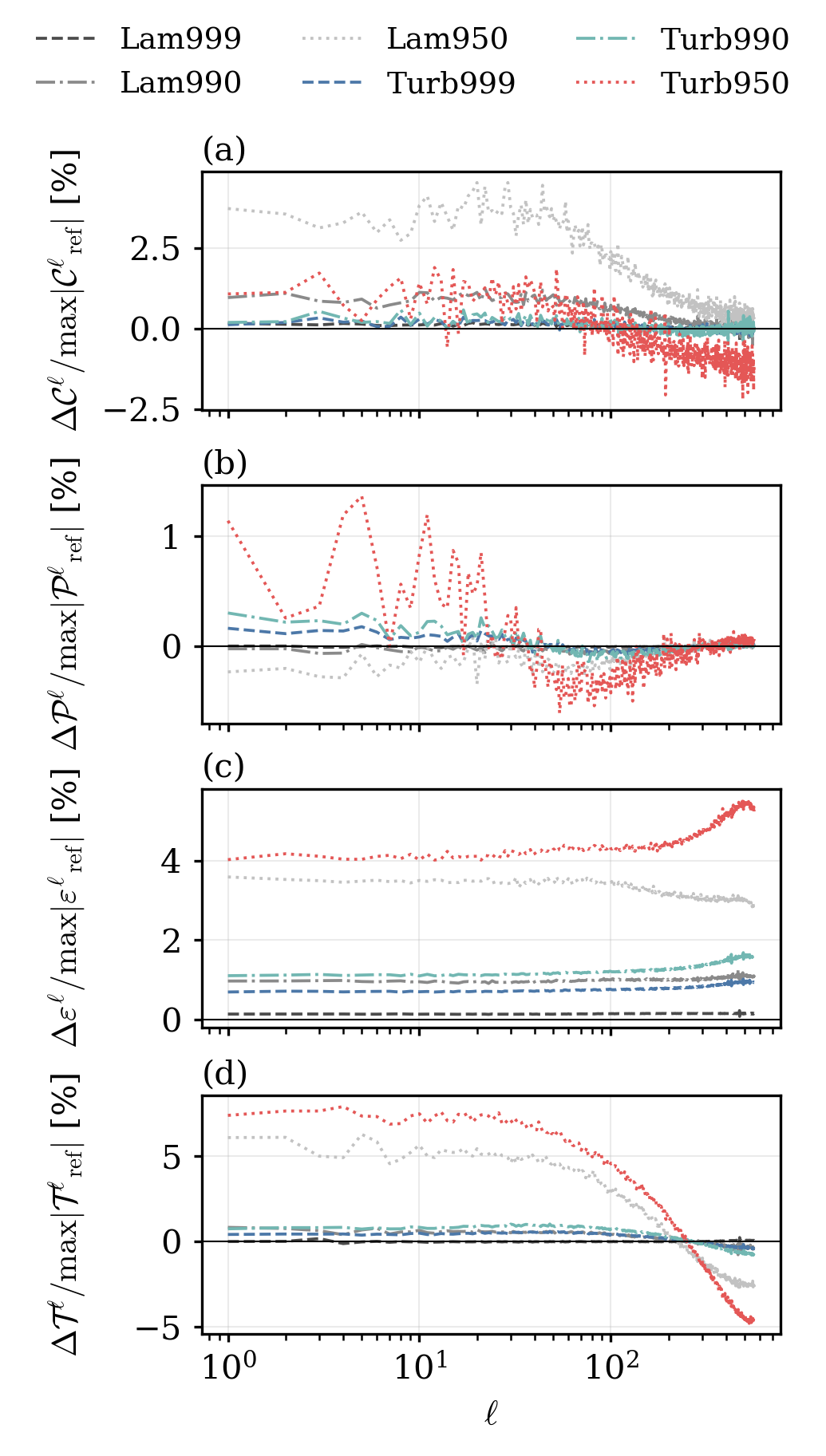}
\caption{Error of the modal budget terms, compressed minus reference normalized by the peak reference value of the same term, against the mode index, for the six compressed databases: (a) convection, (b) production, (c) dissipation, (d) net interscale transfer.}
\label{fig:modes}
\end{figure}

Figure~\ref{fig:modes} extends this view to the whole hierarchy. It shows the error of each budget term, compressed minus reference normalized by the peak of the reference term, against the mode index. The production error stays within $1.4\%$ for every case. The convection error of the $0.950$ targets reaches $4.5\%$ for the laminar and $2.2\%$ for the turbulent inflow, where it changes sign across the hierarchy. The dissipation error is positive at every mode. For the $0.950$ targets, it is $3.6\%$ and $4.0\%$ over the energetic band and grows toward the high-order modes, so the dissipation is overestimated throughout, with the sign discussed below. The transfer error of the $0.950$ targets is positive up to $\ell\approx200$, with $6\%$ and $8\%$ at the first modes, and negative beyond, where the term itself changes sign. These are the ranges built from the discarded scales. 

The right block of table~\ref{tab:errors} lists the relative $L^{2}$ error of each budget term. Every term degrades monotonically as the target is relaxed. All are far more sensitive than the second-order statistics in the left block. The production is practically unaffected even at the $0.950$ target. The gradient-dependent dissipation and viscous diffusion suffer the most. They reach $17\%$ and $10\%$ at the $0.950$ target for both inflows. The convection lies in between, at $5\%$ to $8\%$.

The interscale transfer, built on the third-order moment $\overline{a^{\ell}a^{m}a^{n}}$, separates the fidelity levels most sharply. It stays below $3\%$ at the $0.999$ and $0.990$ targets for both inflows. It rises to $17\%$ for Lam950 and $19\%$ for Turb950. This degradation exceeds any linear term and is almost independent of inflow. The signed sum of $\mathcal{T}^{\ell}$ over the modes vanishes by conservation. The sum of its moduli amounts to $105\%$ and $117\%$ of the total production for the laminar and turbulent references. The error therefore acts on a dominant budget pathway. The retained modes preserve the source of TKE to within $1\%$. The transfer that redistributes it is wrong by an order of magnitude more.

\begin{figure*}
\centering
\includegraphics[width=\textwidth]{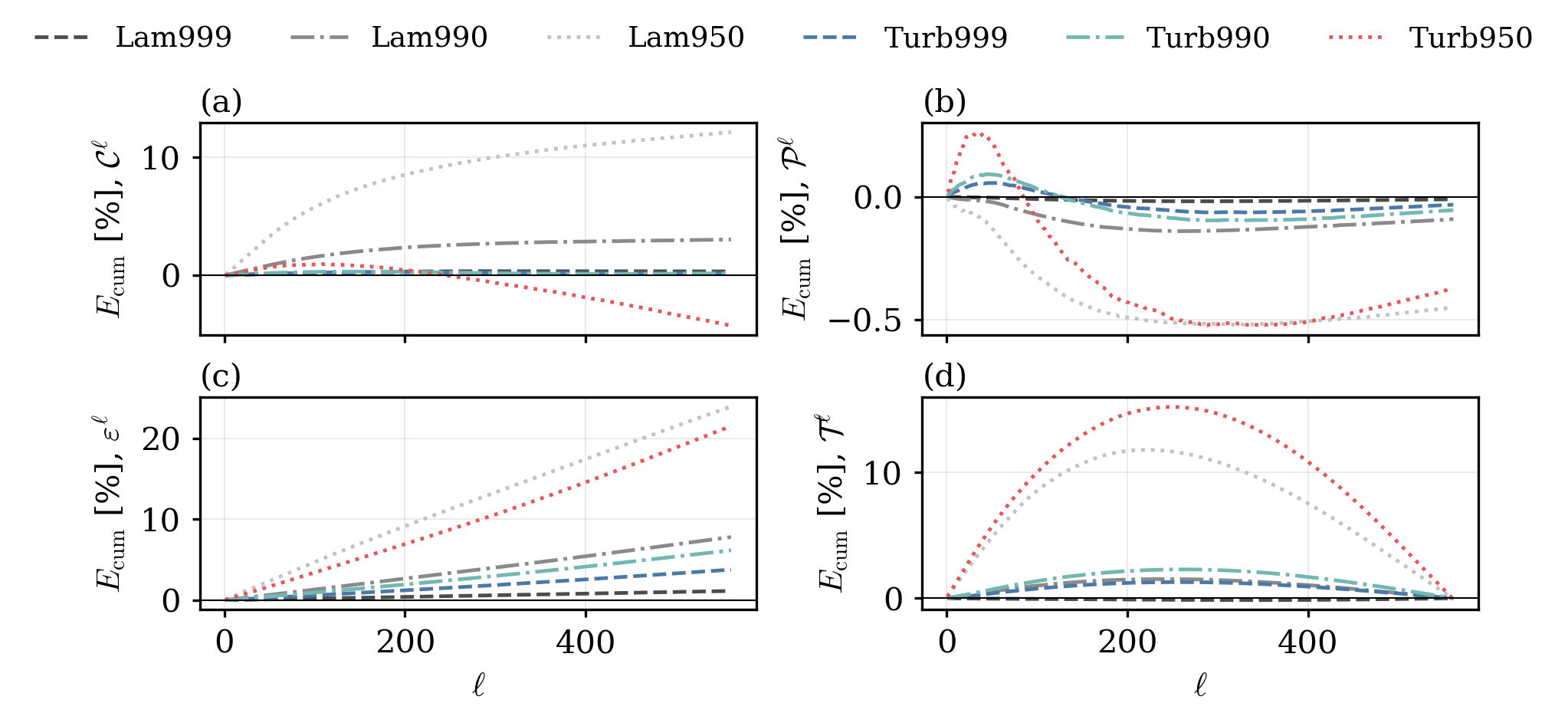}
\caption{Signed cumulative error $E_{\mathrm{cum}}$ of the budget terms against mode index, for the six compressed databases: (a) convection $\mathcal{C}^{\ell}$, (b) production $\mathcal{P}^{\ell}$, (c) dissipation $\varepsilon^{\ell}$, each normalized by the total reference activity of the same term, (d) net interscale transfer $\mathcal{T}^{\ell}$, normalized by the reference production activity. In (a--c) a positive value indicates net over-estimation of the term. In (d) the running sum is the error of the net interscale flux across the mode cut, which vanishes at the last mode by conservation.}
\label{fig:cum}
\end{figure*}

Figure~\ref{fig:cum} resolves these errors mode by mode through the signed cumulative error of eq.~\eqref{eq:ecum}. In panels (a--c), a positive value indicates net over-estimation. The dissipation is systematically over-estimated. Its bias grows almost linearly across the mode hierarchy, to a net $24\%$ for Lam950 and $21\%$ for Turb950. This sign is the opposite of the under-estimation found when dissipation is evaluated on filtered or truncated fields, which lose small-scale gradient content \citep{Olesen2023}. Here both datasets are projected onto matched bases. The over-estimated dissipation is compensated by an under-estimated viscous diffusion (not shown). The discrepancy is a redistribution between the two viscous terms, not a net viscous imbalance. The viscous diffusion is about $2\%$ of the production in magnitude and a pure spatial redistribution, so it is reported in the table only. The production stays unbiased within $0.5\%$ even at the $0.950$ target. The convection bias flips sign with the inflow condition, $12\%$ for the laminar and $-4\%$ for the turbulent inflow at the $0.950$ target. For the interscale transfer (figure~\ref{fig:cum}d), the running sum is divided by the reference production activity. Summed up to mode $\ell$, the transfer is the net rate of energy received by the first $\ell$ modes from the rest of the hierarchy. It is the modal analogue of the spectral flux across a wavenumber cutoff. In a POD basis, the net transfer is directed from low-index to high-index modes \citep{Couplet2003}. The curve is therefore the compression error of the forward interscale flux, which vanishes at the last mode by conservation. Its excursion is positive for every case. The truncation removes the fine-scale recipients and weakens the forward drain of the energetic band. This is the modal counterpart of the energy pile-up produced by sharp spectral truncation \citep{Cichowlas2005}. The excursion grows with the reduction. It peaks near $12\%$ for Lam950 and $15\%$ for Turb950 past the energetic modes.

\subsection{Scaling with the discarded energy}\label{sec:results:scaling}
\begin{figure}
\centering
\includegraphics[width=\columnwidth]{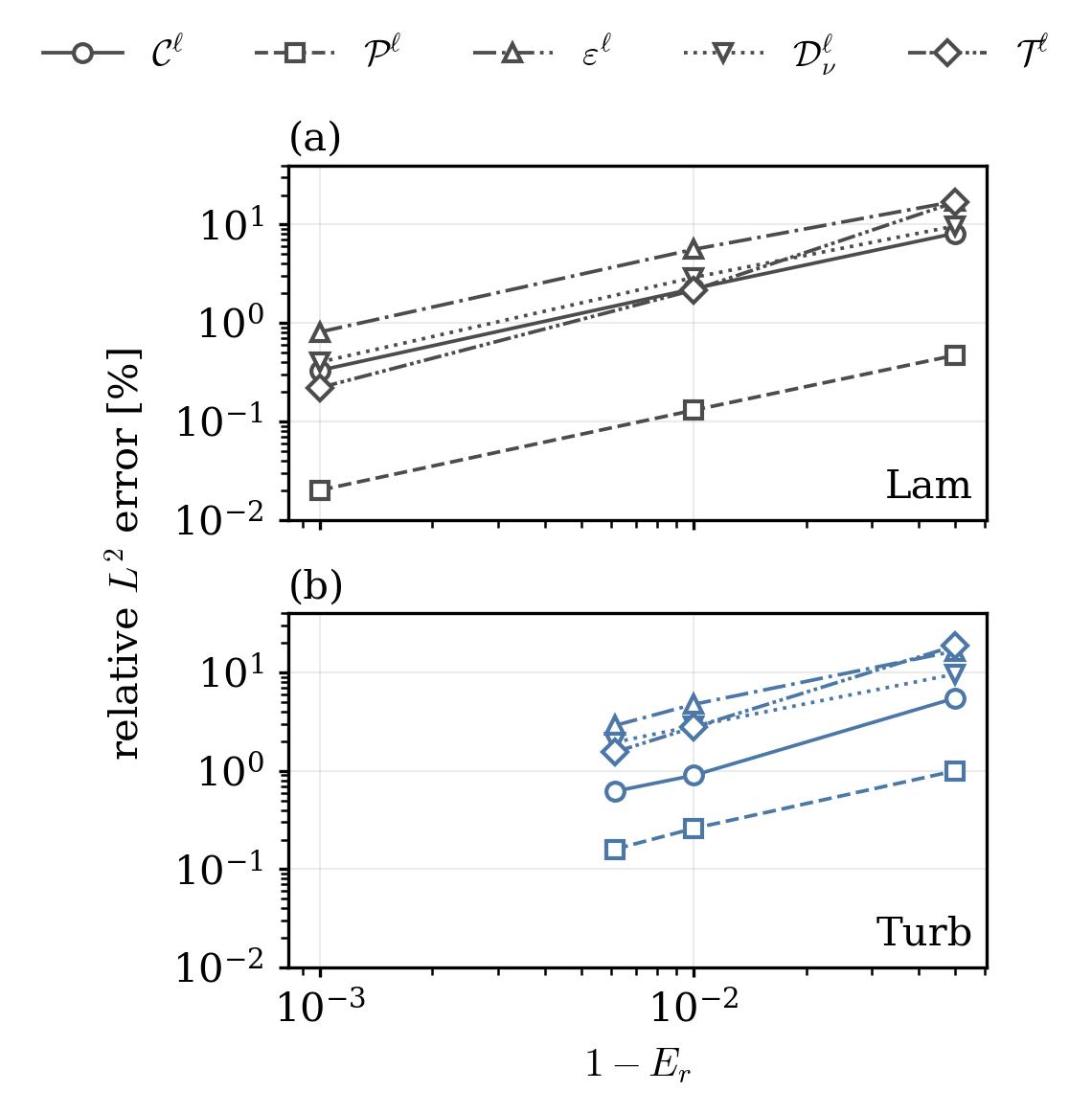}
\caption{Relative $L^{2}$ error of the time-averaged budget terms of table~\ref{tab:errors} against the discarded energy $1-E_r$: (a) laminar inflow, (b) turbulent inflow. The turbulent $0.999$ target is plotted at its achieved energy, $E_r=0.994$.}
\label{fig:scaling}
\end{figure}

The discarded energy $1-E_r$ is the exact relative Frobenius error of the truncated field, eq.~\eqref{eq:corr}. A quantity linear in the field would carry a relative error proportional to $1-E_r$, a power law with exponent one. The budget terms are not linear in the field. The single-mode terms are quadratic in the modal amplitudes and in the gradients of the modes. The transfer is cubic in the amplitudes. How their errors scale with $1-E_r$ is therefore not fixed in advance. Figure~\ref{fig:scaling} plots the relative $L^{2}$ errors of table~\ref{tab:errors} against $1-E_r$ for the two inflows, over the range $10^{-3}\le1-E_r\le5\times10^{-2}$.

Every term follows a power law over this range. The four linear terms scale with exponents between $0.76$ and $0.87$ for both inflows, with one exception: the turbulent convection at $1.06$. The exponents below one mean that the relative error of these terms grows more slowly than the truncation error itself. For the laminar inflow, relaxing the target from $0.999$ to $0.950$ discards fifty times more energy and multiplies the dissipation error by twenty. The terms keep the same order at every target, with the production lowest and the dissipation highest. The transfer scales with an exponent of $1.10$ for the laminar and $1.19$ for the turbulent inflow. It is the least sensitive term at the conservative targets and the most sensitive at the aggressive one. It crosses the dissipation between the $0.990$ and $0.950$ levels. The two inflows fall on offset lines with the same exponents. The turbulent errors are larger at the same discarded energy. The dependence on the retained energy is thus a property of the truncation. The inflow sets the prefactor. 

Two consequences follow. First, the admissible compression level for a given analysis can be read from the truncation error once the prefactor of the flow is known from one calibration run. Second, the transfer error overtakes the dissipation error once the discarded energy exceeds a few per cent. Below that level, the gradient terms set the admissible compression. Above it, the triadic transfer does. For the present databases, the crossover lies between the $0.990$ and $0.950$ targets, which places the conservative compressions of table~\ref{tab:cases}, with factors of $4.5$ to $11$, on the side where every term of the budget is held within $6\%$.

\section{Conclusions}\label{sec:concl}
We have developed an open, distributed, GPU-accelerated low-rank compression framework for large turbulent flow databases and assessed it on two DNS of a controlled diffusion compressor cascade under laminar and turbulent inflow. Rather than certifying the compressed fields by reconstruction error and one-point statistics, we evaluated them through a hierarchy of diagnostics up to the modal turbulent kinetic energy budget and its nonlinear interscale transfer.

The in-plane locations are stacked with the spanwise and temporal coordinates, giving a correlation operator two orders of magnitude larger than that of snapshot POD. The randomized reduction extracts its dominant range at scale. A single decomposition acts on the whole volume rather than local subdomains. The storage ratio is the retained rank over the enlarged column dimension. The energy balance is preserved globally. The budget terms remain assessable.

The central result is that the metrics conventionally used to validate compressed turbulent data are weak where the dynamics live in gradients and correlations. At the aggressive $0.950$ target, the turbulent database is compressed thirteenfold. Every second-order statistic stays within $6\%$. The errors of the dissipation, the viscous diffusion, and the interscale transfer reach $16\%$, $10\%,$, and $19\%$. The production error stays within $1\%$. The compression preserves the energetic source of the budget and corrupts its multiscale redistribution by an order of magnitude more than the one-point measures reveal. Along the blade, the dissipation error of the aggressive target is confined to the viscous sublayer and grows from $17\%$ after reattachment to $23\%$ near the trailing edge of the laminar case and from $13\%$ to $18\%$ along the turbulent one. Over the hierarchy, it is carried by the high-order modes, corrupted by two-thirds, while the energetic modes are held within a few per cent. The signed flux error shows that truncation weakens the forward interscale drain of the energetic scales. For both inflows, the errors of the linear terms scale with the discarded energy to a power of about $0.8$ and that of the transfer to a power above one. The transfer error overtakes the dissipation error once a few per cent of the energy is discarded. The admissible compression level can be read from the truncation error once the prefactor of the flow is known. Conservative retention at $99\%$ to $99.9\%$ compresses the databases by $4.5$ to $11$ times while preserving the entire budget hierarchy, with the interscale transfer below $3\%$.

Compressing one database costs $9.6$ node-hours, $7\%$ of the simulation that produced it. The stored factors are a modal basis of the data. Learned compressors reach larger reduction factors on the same kind of data. None has been certified against a transport budget. Their authors report the loss of the smallest scales on which the dissipation and the triadic transfer are built. A compressed turbulent flow database should be certified against the strictest physics it will be used to study. Here this means the near-wall dissipation and the nonlinear interscale transfer. The admissible level depends on the inflow condition and the intended analysis. Two limitations remain. The truncation minimizes the Frobenius error of the velocity fluctuations, so the retained energy has no meaning for databases that carry variables of incommensurate magnitude. The attainable rank is bounded by the dense reduced eigenproblem on one accelerator, which capped the most conservative turbulent database at $E_r=0.994$.

\section*{CRediT authorship contribution statement}
\textbf{Gustavo Lopes:} Conceptualization, Methodology, Software, Validation, Formal analysis, Investigation, Data curation, Visualization, Writing -- original draft, Funding acquisition. \textbf{Davide Lengani:} Conceptualization, Methodology, Software, Supervision, Project administration, Writing -- review \& editing. \textbf{Dan Henningson:} Supervision, Project administration, Writing -- review \& editing.

\section*{Data availability}
The compression framework is implemented in Python (Dask and CuPy) and is openly archived on Zenodo at \url{https://doi.org/10.5281/zenodo.22013927}, together with a reduced public test case whose input files are archived at \url{https://doi.org/10.5281/zenodo.19481070}. The DNS databases underlying the results are available from the corresponding author on reasonable request.

\section*{Declaration of competing interest}
The authors declare that they have no known competing financial interests or personal relationships that could have appeared to influence the work reported in this paper.

\section*{Acknowledgments}
This project has received funding from the European Union's Horizon Europe research and innovation programme under the Marie Sk{\l}odowska-Curie grant agreement No. 101205669. We acknowledge ISCRA for awarding this project access to the LEONARDO supercomputer, owned by the EuroHPC Joint Undertaking and hosted by CINECA (Italy).

\bibliographystyle{cas-model2-names}
\bibliography{cas-refs}

\end{document}